\documentclass[prl,showpacs,hyperref,superscriptaddress,twocolumn]{revtex4-1}
\usepackage{amsfonts}
\usepackage{amsmath}
\usepackage{amssymb}
\usepackage{graphicx}
\usepackage{subfigure}
\providecommand{\U}[1]{\protect\rule{.1in}{.1in}}
\providecommand{\U}[1]{\protect\rule{.1in}{.1in}}

\begin{document}

\title{Environment-induced sudden transition in quantum discord dynamics}

\author{R. Auccaise}
\affiliation{Empresa Brasileira de Pesquisa Agropecu\'{a}ria, Rua Jardim Bot\^{a}nico 
1024, 22460-000 Rio de Janeiro, Rio de Janeiro, Brazil}

\author{L. C. C\'{e}leri}
\affiliation{Centro de Ci\^{e}ncias Naturais e Humanas, Universidade Federal do ABC, R. 
Santa Ad\'{e}lia 166, 09210-170 Santo Andr\'{e}, S\~{a}o Paulo, Brazil}

\author{D. O. Soares-Pinto}
\affiliation{Instituto de F\'{\i}sica de S\~{a}o Carlos, Universidade de S\~{a}o Paulo, Caixa Postal
 369, 13560-970 S\~{a}o Carlos, S\~{a}o Paulo, Brazil}

\author{E. R. deAzevedo}
\affiliation{Instituto de F\'{\i}sica de S\~{a}o Carlos, Universidade de S\~{a}o Paulo, Caixa Postal
 369, 13560-970 S\~{a}o Carlos, S\~{a}o Paulo, Brazil}

\author{J. Maziero}
\affiliation{Centro de Ci\^{e}ncias Naturais e Humanas, Universidade Federal do ABC, R. 
Santa Ad\'{e}lia 166, 09210-170 Santo Andr\'{e}, S\~{a}o Paulo, Brazil}

\author{A. M. Souza}
\affiliation{Fakult\"{a}t Physik, Technische Universit\"{a}t Dortmund, 44221 Dortmund, Germany}

\author{T. J. Bonagamba}
\affiliation{Instituto de F\'{\i}sica de S\~{a}o Carlos, Universidade de S\~{a}o Paulo, Caixa Postal
 369, 13560-970 S\~{a}o Carlos, S\~{a}o Paulo, Brazil}

\author{R. S. Sarthour}
\affiliation{Centro Brasileiro de Pesquisas F\'{\i}sicas, Rua Dr. Xavier Sigaud 150, 22290-180 
Rio de Janeiro, Rio de Janeiro, Brazil}

\author{I. S. Oliveira}
\affiliation{Centro Brasileiro de Pesquisas F\'{\i}sicas, Rua Dr. Xavier Sigaud 150, 22290-180 
Rio de Janeiro, Rio de Janeiro, Brazil}

\author{R. M. Serra}
\affiliation{Centro de Ci\^{e}ncias Naturais e Humanas, Universidade Federal do ABC, R. 
Santa Ad\'{e}lia 166, 09210-170 Santo Andr\'{e}, S\~{a}o Paulo, Brazil}

\begin{abstract}
Non-classical correlations play a crucial role in the development of quantum
information science. The recent discovery that non-classical correlations can
be present even in separable (unentangled) states has broadened this
scenario. This generalized quantum correlation has been increasing relevance in several fields,
among them quantum communication, quantum computation, quantum phase
transitions, and biological systems. We demonstrate here the occurrence of the sudden-change phenomenon
and immunity against some sources of noise for the quantum discord and its classical counterpart, 
in a  room temperature nuclear magnetic resonance setup. The experiment is performed in a decohering environment causing loss of phase
relations among the energy eigenstates and exchange of energy between system
and environment, resulting in relaxation to a Gibbs ensemble. 

\end{abstract}

\maketitle

The quantum mechanical superposition principle, when applied to composite
systems, foresees the appearance of correlations that cannot be
explained in a classical context \cite{Bell}. Initially, the quantum character of a
correlated system was attributed to the nonlocal aspect of quantum mechanics
and further associated with the violation of Bell's inequalities. The discovery of 
non-separable (entangled) states that do not violate the Bell's theorem led 
eventually to the identification of the quantumness of correlations with the 
separability problem; non-classicality was attributed to entanglement. 
Through the development of quantum information science (QIS),
an operational characterization of entanglement was introduced, as those
correlations that cannot be generated by local operations and classical
communication \cite{Werner}. The development of these ideas 
led to the so-called theory of entanglement, which turned out to be a fruitful branch of
research \cite{Horodeckis}. Entanglement is recognized as an
important resource for several tasks in QIS \cite{Nielsen}. Besides this strong correlation exhibited by entangled states, there is another
kind of non-classical correlation. A composite quantum system in a mixed state
may exhibit some non-classical nature in its correlations even if it is
separable (unentangled) \cite{Ollivier}. Such quantum correlation can be measured as a \textquotedblleft gap\textquotedblright \ between the
quantum and classical versions of mutual information, which is the information-theoretic measure of
the total correlation contained in a bipartite system \cite{Groisman}.

Several approaches have been proposed to quantify this generalized quantum correlation present in a
bipartite mixed quantum state \cite{Ollivier, VlatkoPRL, Termo},
one of which, quantum discord \cite{Ollivier}, has received
special attention \cite{Bradler, Lutz, Lidar, Datta,
Lanyon, Cavalcanti, Maziero1}. Beyond the importance of non-classical correlations (other than entanglement) for the foundations of
quantum mechanics, the relevance of such an issue is increasing in several fields as well in applications in QIS.
Concerning biological systems, it has been suggested that such correlations could play an important role in photosynthesis \cite{Bradler}. In condensed matter physics, quantum correlations of separable states characterize, even at finite temperature, a quantum phase transition
\cite{Maziero1}. In the context of quantum field theory, it has been shown that the Unruh effect may also lead to an abrupt change in the behavior of correlations \cite{Lucas}. For applications in QIS, it is interesting to know that states with nonzero quantum discord cannot be locally broadcast \cite{Piani}, and this kind of non-classicality is also related to a condition for a complete positive map \cite{Lidar}. Quantum discord was proposed as a figure of merit for the quantum  advantage in some computational models without or with little entanglement \cite{Datta,Lanyon}. It also has a relevant role in mixed-state quantum metrology \cite{Metrology}. 

In particular, the quantum discord has been predicted to show peculiar dynamics under decoherence \cite{Sudden}.
Considering two non-interacting qubits (quantum bits) under the action of Pauli maps, it
was shown that, under certain conditions, the decay rate of the correlations
suffers an abrupt modification, a phenomenon denominated sudden change
\cite{Sudden}. Moreover, either the classical or the quantum part of correlation may
be unaffected by decoherence \cite{Sudden, Sudden1}, giving rise to two distinct
decoherence regimes, the classical and the quantum. In the present experiment we demonstrate that the aforementioned phenomena \cite{Sudden, Sudden1} are still present in a \textit{real} thermal environment at room temperature, indicating that such peculiar behavior is quite general. We performed such experimental demonstration in a nuclear magnetic resonance (NMR) setup.

NMR systems have been extensively used as test benches for QIS ideas
\cite{Ivan}. Most of these experiments were performed in
scenarios where the existence of entanglement was ruled out. The quantum
nature of NMR systems at room temperature may be ascribed to quantum correlations of separable
states \cite{Diogo}. The main feature of the technique for QIS is the
excellent control of unitary transformations over the qubit provided by the
use of radio-frequency pulses, which result in unique methods for quantum
state generation and manipulation \cite{Ivan}. In general, for room
temperature liquid state NMR, the Markovian environment in which the qubit is
immersed can be described by two decoherence channels, the amplitude-damping
and the phase-damping, acting locally on each qubit. Our experiment is performed on a liquid state
carbon-13 enriched chloroform sample at room temperature, this sample exhibits two qubits,
encoded in the $^{1}$H and $^{13}$C spin-$1/2$ nuclei. The two-qubit state 
is represented by a density matrix in the high temperature expansion, which takes
the form $\rho_{AB}=\mathbb{I}_{AB}/4+\varepsilon\Delta\rho_{AB}$, where
$\varepsilon=\hbar\omega_{L}/4k_{B}T\sim10^{-5}$ is the ratio between the
magnetic and thermal energies and $\Delta\rho_{AB}$ the deviation
matrix \cite{Chuang, Ivan}. 

Several quantifiers of non-classical correlation have been
proposed. From a thermodynamical approach, Oppenheim and coworkers
have suggested that the non-classical correlations contained in a bipartite
quantum state can be quantified as the quantum deficit, a difference
between the amount of work that can be extracted from a heat bath, by both
and by one of the parts of the composite state \cite{Termo}. In an attempt to
build a unified framework for correlations, Modi and coworkers have proposed a
set of measures for the quantification of their quantum and classical parts,
based on the definition of an entropic distance \cite{VlatkoPRL}. For the NMR context, 
one suitable generalized non-classical correlation may be computed 
from the experimentally accessible variables as a symmetric version of quantum discord
\cite{Symmetric, Diogo} (expanded in the leading order in $\varepsilon$), given by \cite{Diogo}
\begin{equation}
Q(\rho_{AB})=I(\rho_{AB})-\max_{\left\{\Pi_{i}^{A},\Pi_{j}^{B}\right\}}\mathcal{I}(\chi_{AB}), 
\label{QC}
\end{equation}
where the quantum mutual information is given by 
\[
I(\rho_{AB})\approx\frac{\varepsilon^{2}}{\ln2}\left\{2\mbox{tr}(\Delta\rho_{AB}^{2})-\mbox{tr}\left[(\Delta\rho_{A})^{2}\right]-\mbox{tr}\left[(\Delta\rho_{B})^{2}\right]\right\},
\]
and the measurement-induced mutual information is $\mathcal{I}(\chi_{AB})\approx\frac{\varepsilon^{2}}{\ln2}\left\{2\mbox{tr}\left[(\Delta\chi_{AB})^{2}\right]-\mbox{tr}\left[(\Delta\chi_{A})^{2}\right]-\mbox{tr}\left[(\Delta\chi_{B})^{2}\right]\right\}, 
$ with $\chi_{AB}=\mathbb{I}_{AB}/4+\varepsilon\Delta\chi_{AB}$ as the state obtained from $\rho_{AB}$ through a complete projective measurement map ($\Delta\chi_{AB}=\sum_{i,j}\Pi_{i}^{A}\otimes\Pi_{j}^{B}(\Delta\rho_{AB})\Pi_{i}^{A}\otimes\Pi_{j}^{B}$). $\Delta\rho_{A(B)}=\mbox{tr}_{B(A)}\{\Delta\rho_{AB}\}$ is the reduced deviation matrix while $\Delta\chi_{A(B)}$ stands for the reduced measured deviation matrix in the subspace $A(B)$. The classical counterpart of Eq. (\ref{QC}) is $C(\rho_{AB})=\max_{\left\{\Pi_{i}^{A},\Pi_{j}^{B}\right\}}\mathcal{I}(\chi_{AB})$. It is worth mentioning that $Q(\rho_{AB})=0$ if and only if $\rho_{AB}$ can be cast in terms of local orthogonal basis. In other words, the symmetric quantum discord is zero if and only if $\rho_{AB}$ has only classical correlations or no correlations at all \cite{Symmetric}. The aforementioned symmetric correlation quantifiers can be computed directly from the experimentally reconstructed deviation matrix.
 
Let us consider the class of states with maximally mixed marginals
($\rho_{A(B)}=\mathbb{I}_{A\left(  B\right)  }/2$), also known as Bell-diagonal
states:
\begin{equation}
\rho_{AB}=\frac{1}{4}\left(  \mathbb{I}_{AB}+\sum_{i=x,y,z}c_{i}\sigma_{i}%
^{A}\otimes\sigma_{i}^{B}\right)  , \label{State}%
\end{equation}
where $\sigma_{i}^{A(B)}$ is the $i$th component of the standard Pauli operator
acting on the $A$ ($B$) subspace, $c_{i}=\mbox{tr}\left[  \rho_{AB}\sigma
_{i}^{A}\otimes\sigma_{i}^{B}\right]  $, and $\mathbb{I}_{AB}$ is the identity
operator. It was theoretically predicted \cite{Sudden} that, depending on the geometry (encoded in the relations between the parameters
$c_{i}$) of the state (\ref{State}), the evolution of the state correlations
under the action of Pauli channels presents a peculiar behavior \cite{Sudden}.
Considering the phase-damping channel, if the state components are related as follows: $\left\vert c_{z}\right\vert \geq\left\vert c_{x}%
\right\vert ,\left\vert c_{y}\right\vert $, it was shown the classical correlation is not
affected by decoherence, while the quantum correlation decays exponentially.
On the other hand, if $\left\vert c_{x}\right\vert \geq\left\vert
c_{y}\right\vert ,\left\vert c_{z}\right\vert \ $or $\left\vert c_{y}%
\right\vert \geq\left\vert c_{x}\right\vert ,\left\vert c_{z}\right\vert $,
and $\left\vert c_{z}\right\vert \neq0$, the correlations
exhibit the so-called sudden change in behavior \cite{Sudden}. The classical correlation
decays exponentially until a specific moment in time, thereafter remaining
constant, while the quantum correlation suddenly changes its decay rate
at that moment. For the other two Pauli channels, we can see the same
behavior, except that the relations between the three components of state
(\ref{State}) are exchanged \cite{Sudden}. 

\begin{figure}[tbh]
\centering
\includegraphics[scale=0.8]{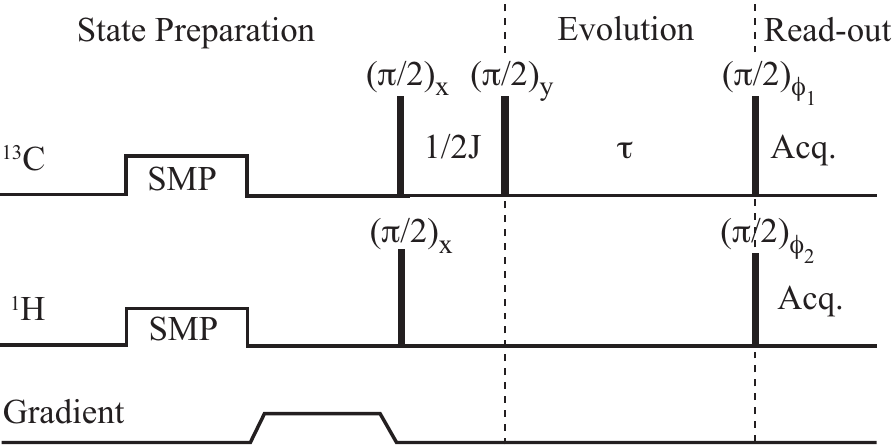}\caption{Sketch of the pulse sequence used experimentally to follow the
dynamics of quantum and classical correlations under decoherence. The sequence
consists of three blocks: the initial state preparation,
relaxation delay and read-out by quantum state tomography.}
\label{pulseq}
\end{figure}

In order to demonstrate such non-trivial dynamics we prepare two initial Bell-diagonal states with specific relations between its components. This involves the experimental mapping of the second term of the right hand side of
Eq. (\ref{State}) onto $\Delta\rho_{AB}$. Starting from the thermal equilibrium state, the mapping consists of three main steps, in which the following are applied to the sample: ($i$) a strongly modulated pulse,
($ii$) a magnetic field gradient and ($iii$) a pseudo Einstein-Podolsky-Rosen gate implemented by radio-frequency pulses \cite{Chuang, Ivan}. After the preparation of the state, it is left to evolve freely under decoherence and the dynamics of the system is followed by quantum state tomography \cite{long2001}. Figure \ref{pulseq} illustrates the pulse sequence in the experimental procedure \cite{supplementary}.

\begin{figure}[tbh]
\includegraphics[scale=0.35]{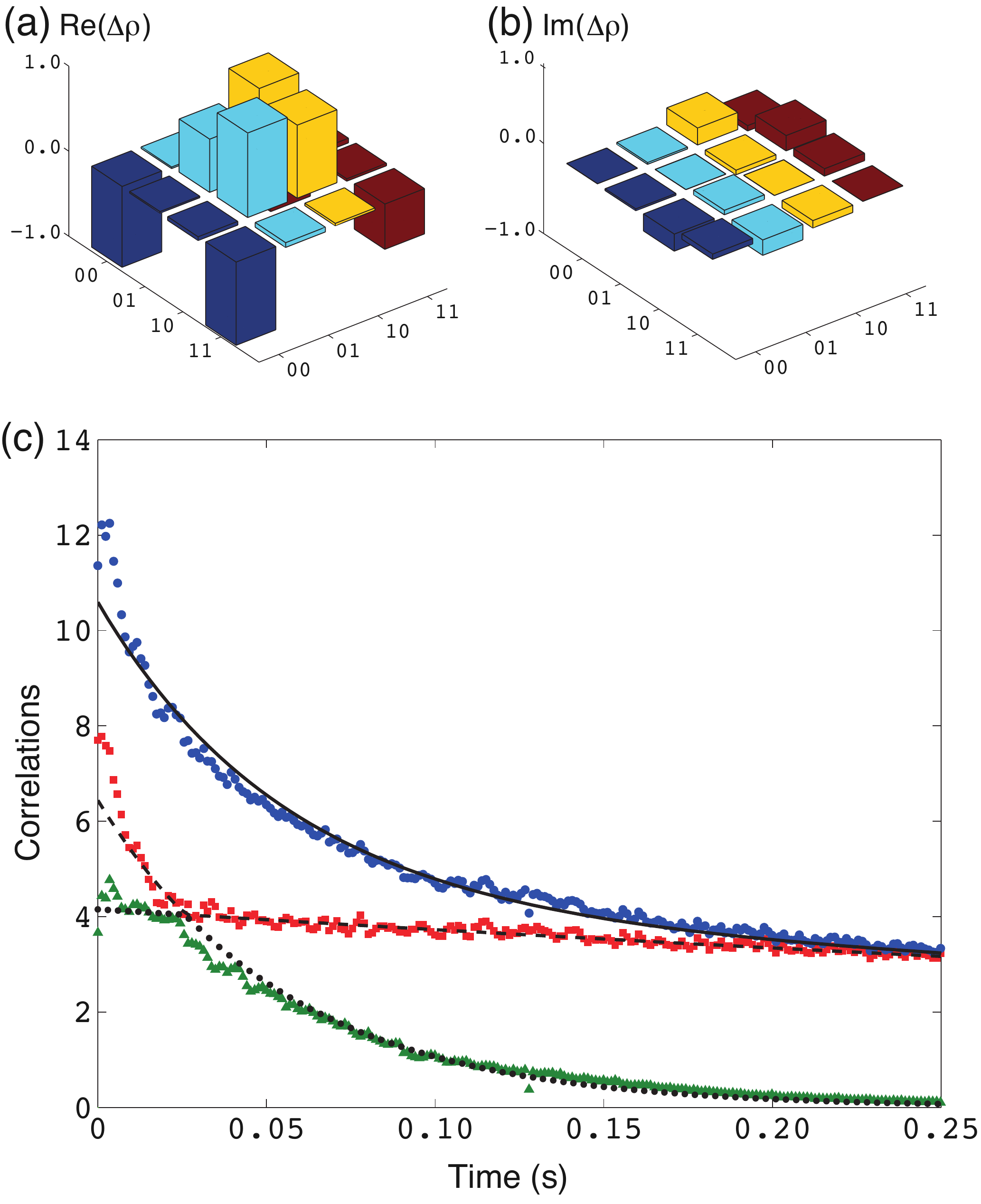}
\caption{(Color online) Sudden change in behavior of correlations. (a) bar representation of the real and, (b) imaginary parts of the initial deviation matrix for the sudden-change experiment, reconstructed  by quantum state tomography. We adopted the usual computational basis, where $\left| 0 \right\rangle$ and $\left| 1 \right\rangle$ represents the eigenstates of $\sigma_{z}$ for each qubit. (c) displays the predicted
sudden change in behavior of the correlations during their dynamic evolution
to thermal equilibrium. The blue circles are the experimental data for
the quantum mutual information, while the red squares and green triangles
represent the classical and quantum correlations, respectively. The black lines are the theoretical predictions. The initial state is analogous to the state in Eq. (\ref{State}) with $\left|c_{x}\right|, \left|c_{y}\right| > \left|c_{z}\right|$. The correlations are displayed in units of $(\varepsilon^{2}/\ln 2)$bit.}
\label{Fig2}
\end{figure}

\begin{figure}[tbh]
\includegraphics[scale=0.35]{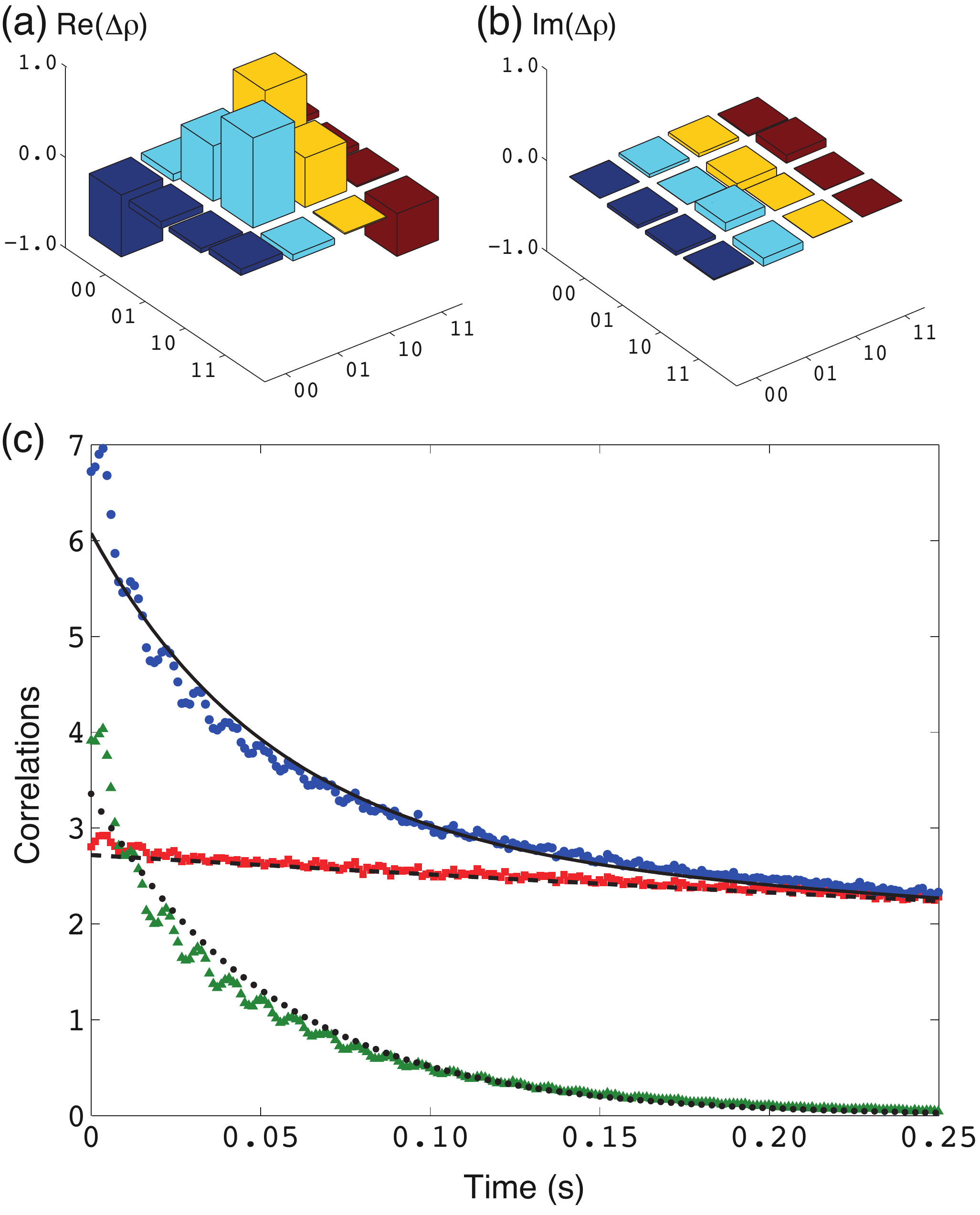}
\caption{(Color online) Immunity against decoherence. (a) bar representation of the real and, (b) imaginary parts of the initial deviation matrix, reconstructed by quantum state tomography. (c)
shows that the classical correlation is not affected by the
action of the phase-damping channel. The small decay rate is entirely due to
the amplitude-damping channel. The displayed pattern is the
same as in Fig. 2. For this experiment, the initial state is analogous to state in Eq. 
(\ref{State}) with $\left|c_{z}\right| > \left|c_{x}\right|, \left|c_{y}\right|$. 
The correlations are computed in units of $(\varepsilon^{2}/\ln 2)$bit.}
\label{Fig3}
\end{figure}

The relaxation process that causes phase decoherence and energy dissipation is due,
mainly, to internal molecular or atomic motions that cause random fluctuations
in the electromagnetic field in which the qubits are immersed. These
fluctuations are characterized by the spectral densities that encode features of the motion, such as geometry and correlation times. The environment can be modeled by two independent quantum channels, described as the generalized
amplitude-damping and the phase-damping channels. In previous
investigations of sudden-change dynamics \cite{Sudden, Sudden1,
QDexp}, the energy exchange channel was not take into account, however in the present
experiment this relaxation source is unavoidable. It is remarkable  that the sudden-change dynamics is still happen even in the presence of a  thermal environment as can be seen in Fig. \ref{Fig2}. The small
decay rate of the classical correlation, after the transition point (where it could be constant under a phase damping), is due to
the presence of the thermal noise source. The theoretical predictions presented in Fig. \ref{Fig2} are in good agreement with
the experimental data \cite{supplementary}. We observe a sudden transition
between two decoherence regimes \cite{Sudden}, i.e., classical and quantum
decoherence \cite{Sudden1}. During the first regime ($t\lessapprox0.027$ s) the
decoherence affects more strongly the classical aspects of correlation,
leaving the quantum aspects with a small decay. After the sudden-change point
($t\gtrapprox0.027$ s) the classical correlation becomes more robust
against decoherence and the noise degrades more effectively the quantumness of
correlations. This confirms that the phenomena predicted in \cite{Sudden,Sudden1} for
phase environments takes place even in the presence of an additional thermal noise.

Figure \ref{Fig3} also confirms the theoretical prediction \cite{Sudden} that, in
some states (depending on their geometry), the classical correlation may be robust against phase noise. The small decay rate of this
correlation plotted in Fig. \ref{Fig3} comes entirely from the thermal source. Once more, the theoretical curves presents
very good agreement with the experimental data.

In both of the Figs. 2(c) and 3(c), the theoretical data analysis guarantees
that the small decay rate presented by the classical correlation, where it
should vanish, comes exclusively from the action of the thermal channel on
the system. It is worth mentioning that the oscillations shown by the
experimental curves in both figures are due to experimental details (discussed
in \cite{supplementary}).

Correlations are ubiquitous in nature. The discovery that separable quantum states can exhibit non-classical
correlation other than entanglement has led to a new understanding of the
quantum aspects of a physical system. Despite the relatively great number of theoretical articles
concerning decoherence effects, up to date, only few experimental investigations of the correlation
dynamics have been reported previously in literature \cite{QDexp,Diogo,Witness}. The predictions of refs. \cite{Sudden, Sudden1} were tested in a \textit{simulated} phase noise environment \cite{QDexp}. This experiment was performed in an optical setup, in which the action of a phase-damping channel was \textit{simulated} in a \textit{controllable} way by a birefringent medium and the qubits were encoded in photon polarizations
\cite{QDexp}. Under the action of a real global environment, the dynamics of correlations was investigated in a NMR quadrupolar system \cite{Diogo}.

In the present experiment we observed the environment-induced sudden change phenomena in a \textit{real} (uncontrollable) thermal noise environment at room temperature. The environment-induced sudden change takes place in the course of the relaxation of two nuclear spins to the Gibbs state. This demonstrate that this phenomena may occur in a general context when a non-equilibrium system relax to a thermal equilibrium state. The methods employed in our experiment to follow the dynamics of quantum discord and its classical counterpart may be applied to other molecules (including biological ones) and also to other experimental contexts where high mixed states are present. The two decoherence regimes observed may have important consequences for applications in QIS, since the nature of correlations in a given mixed state system is somehow related to the quantum advantage for performing a given task (as for example in quantum metrology \cite{Metrology}) or preventing local broadcasting of information \cite{Piani}. 

\begin{acknowledgements} 
The authors acknowledge financial support from UFABC, CNPq, CAPES, FAPESP, and FAPERJ. This work was performed as part of the Brazilian National Institute of Science and Technology for Quantum Information (INCT-IQ).
\end{acknowledgements} 

\appendix

\section{\textbf{Supplementary information - Details on experimental and calculation procedures}}

\subsection{\textbf{NMR experiments}}

NMR experiments were performed on a two-quantum bit system composed of
nuclear spins of $^{1}$H and $^{13}$C atoms in carbon-13 enriched chloroform (CHCl$_{3}$). The sample was prepared by mixing $100$ mg of $99$ \%
$^{13}$C-labelled CHCl$_{3}$ in $0.2$ mL of $99.8\%$ CDCl$_{3}$ in a $5$ mm
NMR tube (both compounds provided by the Cambridge Isotope Laboratories Inc.). NMR experiments were carried out at $25^{\circ}$C in a Varian $500$
MHz Infinity Plus spectrometer located at the Brazilian Center for Physics
Research (CBPF, Rio de Janeiro). A Varian $5$ mm double resonance probehead equipped with a
magnetic field gradient coil was used. $\pi/2$ pulses of length $7.4$ $\mu$s
and $8.0$ $\mu$s were applied to $^{1}$H and $^{13}$C, respectively. For
specific state preparation, a five-steps strongly modulated pulses (SMP)
\cite{fortunato2002} was designed using a MATLAB self-written routine based on
a SIMPLEX optimization protocol. Spin-lattice relaxation times (T$_{1}$) for
$^{1}$H and $^{13}$C nuclei, measured by the inversion recovery pulse
sequence, were $2.5$ s and $7$ s, respectively. Spin-spin relaxation times
(T$_{2}$), measured by a CPMG pulse sequence, were estimated to be about $1.8$ s
and $0.29$ s for $^{1}$H and $^{13}$C nuclei, respectively. The recycle delay
(d$_{1}$) was set at $40$ s in all experiments. As no refocusing
pulse was used, the effective transversal relaxation times are given by $T_{2}^{\ast
}=0.31$ s and $T_{2}^{\ast}=0.12$ s for $^{1}$H and $^{13}$C nuclei, respectively.

The nuclear spin Hamiltonian accounting for the relevant NMR interactions is
given by

\begin{widetext}
\begin{eqnarray}
\mathcal{H} &=&-\left( \omega _{H}-\omega _{rf}^{H}\right) \mathbf{I}%
_{z}^{H}-\left( \omega _{C}-\omega _{rf}^{C}\right) \mathbf{I}_{z}^{C}+2\pi
J\mathbf{I}_{z}^{H}\mathbf{I}_{z}^{C} \nonumber \\
&&+\omega _{1}^{H}\left( \mathbf{I}_{x}^{H}\cos \varphi ^{H}+\mathbf{I}_{y}^{H}\sin \varphi
^{H}\right) +\omega _{1}^{C}\left( \mathbf{I}_{x}^{C}\cos \varphi ^{C}+\mathbf{I}
_{y}^{C}\sin \varphi ^{C}\right),
\label{Hamiltoniano}
\end{eqnarray}
\end{widetext}
where $\mathbf{I}_{\alpha}^{H}\left(  \mathbf{I}_{\beta}%
^{C}\right)  $ is the spin angular momentum operator in the $\alpha
,\beta=x,y,z$ direction for $^{1}$H ($^{13}$C); $\varphi^{H}\left(
\varphi^{C}\right)  $ defines the direction of the RF field (pulse phase) and
$\omega_{1}^{H}\left(  \omega_{1}^{C}\right)  $ is the RF nutation frequency
(RF power) for $^{1}$H ($^{13}$C) nuclei. The first two terms describe the Zeeman
interaction between the $^{1}$H and $^{13}$C nuclear spins and the main
magnetic field B$_{0}$ . The corresponding frequencies are $\omega_{H}%
/2\pi\approx500$ MHz and $\omega_{C}/2\pi\approx125$ MHz. The third term is
due to a scalar spin-spin coupling of $J\approx215.1$ Hz. The fourth and
fifth terms represent the radio-frequency field (RF) that may be applied to the
$^{1}$H and $^{13}$C nuclear spins, respectively. Besides the interactions written
above, there is a time-dependent coupling of the nuclear spins with the
environment that includes all fluctuating NMR interactions, which accounts for
the spin relaxation, e.g., $^{1}$H - $^{13}$C dipolar spin-spin couplings,
interactions with the chlorine nuclei, etc.

To follow the decoherence of the correlations and distinguish the distinct
regimes described in the main text, an initial state such as that shown in Eq.
(1) of the the main text needs to be prepared with specific $c_{i}$'s . Therefore, from the
experimental perspective, it is essential to prepare a state in which populations
and coherences of the deviation matrix can be set accordingly. To that end,
a diagonal state was created from the thermal equilibrium by a $4-8$ ms SMP
followed by a $2$ ms magnetic field gradient pulse. This produced a
deviation matrix as shown in Eq. (\ref{estadodiagonal}), where
the elements $\alpha$, $\beta$, $\gamma$ and $\delta
$ are defined by the optimized SMP pulse.
Next, this diagonal deviation matrix is transformed into an X-type
matrix, such as that in Eq. (\ref{estadoX}), by a pseudo-EPR gate
implemented by the RF pulse sequence described in reference
\cite{chuang1998PRSLA}.

\begin{align}
\Delta\rho &  =\left[
\begin{array}
[c]{cccc}%
\alpha & 0 & 0 & 0\\
0 & \beta & 0 & 0\\
0 & 0 & \gamma & 0\\
0 & 0 & 0 & \delta
\end{array}
\right]  \text{,}\label{estadodiagonal}\\
\Delta\rho &  =\frac{1}{2}\left[
\begin{array}
[c]{cccc}%
\alpha+\gamma & 0 & 0 & -\alpha+\gamma\\
0 & \beta+\delta & -\beta+\delta & 0\\
0 & -\beta+\delta & \beta+\delta & 0\\
-\alpha+\gamma & 0 & 0 & \alpha+\gamma
\end{array}
\right]  \text{.} \label{estadoX}%
\end{align}

The pulse sequence used in the NMR experiments is illustrated in Fig.
1 of the main text. The SMP was designed to produce a diagonal state with
populations $\alpha$, $\beta$, $\gamma$ and $\delta$, so that, after the
transformation by the pseudo-EPR gate, a state with a deviation matrix in the form equivalent to 
the state in Eq. (2) of the main text, with specified $c_{i}$'s, was obtained. After this
state preparation step, the system was left to evolve for a period
$\tau=m/4J$, with $m=1,2,3,4,..$. During this evolution period (considering on-resonance evolution), two terms of the Hamiltonian shown in Eq. (\ref{Hamiltoniano}) act on the nuclear spins: the scalar coupling, which
produces oscillations of the quantum coherence, and the time-dependent term,
which accounts for the system relaxation. It is worth mentioning that the elements 
in the main and cross diagonals of the deviation matrix account for populations and 
quantum coherences that do not oscillate with the $J$ coupling evolution, so that, for a perfect 
X-type deviation matrix, the aforementioned oscillation should not be observed. However, 
the experimental deviation matrices have small coherences off the X positions, which 
oscillate during the evolution. This leads to small oscillations in the spectral line intensities 
used to perform the state tomography and, because the tomography process involves solving 
coupled equations, this oscillatory behaviour is captured in the main and cross diagonal 
elements, causing the oscillations observed in the correlations shown in Figs. 2 (c) and 3 (c) of main text. 
The oscillatory evolution may be suppressed from the experimental data by setting the evolution step to a
multiple of $1/J$. However, in order to show unequivocally the sudden change
in the quantum correlations, the evolution step was set to $1/4J$,
giving rise to the small oscillations observed in the evolution of the
correlations depicted in Figs. 2 (c) and 3 (c) of the main text. After the evolution period, the quantum state was read-out by quantum state tomography, as described in reference \cite{long2001,teles}. Multi-step increments of
$\tau$ in successive experiments allowed the effect of the spin
environment on the initial deviation matrix to be followed. Final states
corresponding to $251$ distinct $\tau$ values were acquired. The
high homogeneity of the static magnetic field was guaranteed by the linewidth of
$\approx0.95$ Hz in $^{1}$H spectra.

\subsection{\textbf{Decoherence analysis}}

The decoherence process is theoretically analysed through the operator sum representation
technique \cite{Nielsen}, in which the evolution of the density operator is
given by
\[
\rho\left(  t\right)  =\sum_{k} E_{k}\left(  t\right)  \rho\left(  0\right)
E_{k}^{\dagger}\left(  t\right)  ,
\]
where $E_{k}\left(  t\right)  $ are the well-known Kraus operators.

\textit{Amplitude-Damping Channel}: In this scenario, the amplitude-damping
channel is described by the following set of operators
\begin{subequations}
\begin{align*}
E_{0}  &  =\sqrt{\gamma}%
\begin{pmatrix}
1 & 0\\
0 & \sqrt{1-p}%
\end{pmatrix}
,\quad E_{1}=\sqrt{\gamma}%
\begin{pmatrix}
0 & \sqrt{p}\\
0 & 0
\end{pmatrix}
,\\
E_{2}  &  =\sqrt{1-\gamma}%
\begin{pmatrix}
\sqrt{1-p} & 0\\
0 & 1
\end{pmatrix}
,\quad E_{3}=\sqrt{1-\gamma}%
\begin{pmatrix}
0 & 0\\
\sqrt{p} & 0
\end{pmatrix}
,
\end{align*}
\end{subequations}
where, in the NMR context, $\gamma=1/2-\varepsilon/2$ and $p=1-\exp\left(
-t/T_{1}\right)  $, $T_{1}$ being the longitudinal relaxation time of the
qubit under consideration. We observe that in our case, the relaxation times are
different for the two qubits, since they have distinct Larmor frequencies.

\textit{Phase-Damping Channel}: For this case we have

\begin{subequations}
\[
E_{0}=\sqrt{1-\frac{\lambda}{2}}%
\begin{pmatrix}
1 & 0\\
0 & 1
\end{pmatrix}
,\quad E_{1}=\sqrt{\frac{\lambda}{2}}%
\begin{pmatrix}
1 & 0\\
0 & -1
\end{pmatrix}
,
\]
\end{subequations}
where $\lambda=1-\exp\left(  -t/T_{2}\right)  $ with $T_{2}$ being the
transverse relaxation time associated with the qubit.

The agreement between the model and the experimental data, as observed 
in Figs. 2 (c) and 3 (c) of the main text, is fairly good. In the theoretical model we neglected a number of possible error sources, such as magnetic field inhomogeneities or oscillations of the residual off-diagonal elements of the deviation matrix. In spite of that, all the singular features predicted by the quantum model are very clear in the experimental data. 

Within this model and taking into account that both environments act
independently on each qubit, it is straightforward to eliminate from the
experimental data the contribution of the amplitude-damping channel to the
dynamics of the correlations. This procedure removes the small decay rate
of the classical correlation (which it should be constant) in both Figs. 2 (c)
and 3 (c) of the main text, proving that this correlation is really immune to the phase-damping channel in certain cases, corroborating the theoretical predictions made in ref. \cite{Sudden}.

\end{document}